\documentclass{osa-article}

\journal{oe}

\articletype{Research Article}

\usepackage{lineno}
\usepackage{ulem}
\usepackage{xcolor}
\usepackage{upgreek}
\usepackage{notes2bib}
\usepackage{soul}  

\begin{document}

\title{Intra-Cavity Frequency-Doubled VECSEL System for Narrow Linewidth Rydberg EIT Spectroscopy}
\author{Joshua C. Hill\authormark{1,*},
William K. Holland\authormark{1},
Paul D. Kunz\authormark{1},
Kevin C. Cox\authormark{1},
Jussi-Pekka Penttinen\authormark{2,3},
Emmi Kantola\authormark{2}, and
David H. Meyer\authormark{1}}

\address{
\authormark{1}DEVCOM Army Research Laboratory, 2800 Powder Mill Road, Adelphi, Maryland 20783, USA\\
\authormark{2}Vexlum Ltd, Korkeakoulunkatu 3, 33720 Tampere, Finland\\
\authormark{3}Optoelectronics Research Centre, Tampere University, 33720 Tampere, Finland\\
}

\email{\authormark{*}joshua.c.hill49.civ@army.mil}

\begin{abstract}
High-power, narrow-linewidth light sources in the visible and UV spectra are in growing demand, particularly as quantum information and sensing research proliferates. Vertical external-cavity surface-emitting lasers (VECSELs) with intra-cavity frequency conversion are emerging as an attractive platform to fill these needs. Using such a device, we demonstrate 3.5\,MHz full-width half-maximum Rydberg-state spectroscopy via electromagnetically induced transparency (EIT). The laser's 690\,mW of output power at a wavelength of 475\,nm enables large Rabi frequencies and strong signal-to-noise ratio in shorter measurement times. In addition, we characterize the frequency stability of the VECSEL using the delayed self-heterodyne technique and direct comparison with a commercial external-cavity diode laser (ECDL). We measure the pre-doubled light's Lorentzian linewidth to be $2\pi\times5.3(2)$\,kHz, and the total linewidth to be $2\pi\times23(2)$\,kHz. These measurements provide evidence that intra-cavity frequency-doubled VECSELs can perform precision spectroscopy at and below the MHz level, and are a promising tool for contemporary, and future, quantum technologies.
\end{abstract}


\section{Introduction} 
Many modern quantum science applications demand continuous wave (CW) laser power in excess of one watt at wavelengths ranging from the deep ultraviolet (UV) to infrared. Moreover, such applications often simultaneously require laser linewidths narrower than the linewidth of the particular energy transition being manipulated. Laser cooling and trapping \cite{phillipsLaserCoolingElectromagnetic1985,kanedaContinuouswaveSinglefrequency2292016}, large-momentum-transfer atom interferometry \cite{noursharghCirculatingPulseCavity2021}, high-fidelity Raman-mediated qubit gates \cite{monroeProgrammableQuantumSimulations2021}, and Rydberg-atom electrometry \cite{meyerOptimalAtomicQuantum2021,meyerWaveguideCoupledRydbergSpectrum2021} are examples of topics attracting considerable interest that would benefit from the availability of lower-cost, higher-power lasers with satisfactory spectral characteristics at these wavelengths \cite{burdOpticallyPumpedSemiconductor2015}. 
 
Presently, the range of needs for such applications is met with complex systems including titanium-sapphire and amplified external-cavity diode (ECDL) lasers in combination with sum-frequency and second-harmonic generation. Popular techniques for obtaining increased laser power include chains of multiple lower-power sources in an injection locking scheme, or tapered amplifiers that often suffer from poor quality output intensity profiles and amplification of spontaneous emission. Complexity generally increases cost, decreases stability, and can limit the range of potential uses.

Vertical external-cavity surface-emitting lasers (VECSEL), also referred to as optically-pumped semiconductor lasers, are an established approach for generating high-power CW laser light \cite{kuznetsovVECSELSemiconductorLasers2010}.
Developments over the past decade have led to increasing interest in atomic physics applications due to the availability of gain chips with sufficient performance spanning infrared wavelengths in a simple form factor \cite{burdOpticallyPumpedSemiconductor2015,burdSinglefrequency571nmVECSEL2016,burdVECSELSystemsGeneration2016, burdVECSELSystemsQuantum2020,paboeufFrequencyStabilizationUltraviolet2013,tinsleyWattlevelBlueLight2021,paulDopplerfreeSpectroscopyMercury2011}. When combined with frequency-conversion, the VECSEL architecture is capable of producing watts of single-mode optical power in the deep UV to visible portions of the spectrum \cite{zhang23wattSinglefrequencyVerticalexternalcavity2014,tinsleyWattlevelBlueLight2021,rantamakiWatt56Um2013,laurainMultiwattpowerHighlycoherentCompact2010,rakoHighpowerSinglefrequencyIntracavity2020,chillaHighpowerOpticallyPumped2004,heinEfficient460nmSecondHarmonic2011}. By incorporating the nonlinear medium into the lasing cavity, simple and efficient generation of second-harmonic light can be realized \cite{guinaOpticallyPumpedVECSELs2017}.Despite these potential benefits, use of intra-cavity frequency doubled VECSELs for precision atomic spectroscopy has been limited. To our knowledge, such lasers have been used to observe spectroscopic features with widths only down to approximately 10\,MHz in, for example, experiments exciting molecular iodine \cite{burdVECSELSystemsGeneration2016, burdSinglefrequency571nmVECSEL2016}. Intra-cavity doubling has been used with VECSELs to perform spectroscopy of neutral cadmium at the 90\,MHz level \cite{tinsleyWattlevelBlueLight2021}, and extra-cavity quadrupled systems have been used to excite trapped beryllium ions \cite{burdVECSELSystemsQuantum2020} and mercury vapors \cite{paulDopplerfreeSpectroscopyMercury2011}.

In this work we present an analysis of the suitability of a high-power intra-cavity doubled VECSEL for narrow atomic spectroscopy at the sub-100\,kHz level by investigating the device's spectral properties using the delayed self-heterodyne technique and an analysis of long term stability and low-frequency noise via direct comparison with a reference ECDL system. These results are complemented by measurements of Rydberg-atom electromagnetically induced transparency (EIT) in a room-temperature rubidium vapor with the VECSEL performing one segment of the two-photon coherent excitation \cite{mohapatraCoherentOpticalDetection2007}. We observe linewidths as narrow as 3.5\,MHz full width half maximum (FWHM) in the EIT probe beam's transmission. This spectroscopic measurement represents, to our knowledge, the narrowest feature resolved using an intra-cavity frequency-doubled VECSEL.This analysis and demonstration highlights the suitability of the VECSEL platform for many demanding quantum science applications.

\section{VECSEL description}

Vertical external-cavity surface-emitting lasers have a long history of generating high-power CW laser light over a wide range of infrared wavelengths \cite{kuznetsovVECSELSemiconductorLasers2010, guinaOpticallyPumpedVECSELs2017}.
Combined with efficient frequency doubling via nonlinear crystals, VECSELs are a promising technology for emission in the UV to visible portions of the spectrum. Advantageous properties, including a lack of both relaxation oscillations and amplified spontaneous emission, can allow VECSELs to simultaneously achieve high output powers and frequency noise characteristics comparable, or potentially superior, to those of ubiquitous ECDL systems \cite{burdVECSELSystemsGeneration2016,moriyaSubkHzlinewidthVECSELsCold2020}.
 
The VECSEL's external cavity, which distinguishes it from a vertical cavity surface emitting laser (VCSEL), allows for \textit{intra-cavity} frequency conversion. Incorporating the nonlinear optical element within the laser cavity can make the device more compact without the need for an additional cavity and its associated components. However, the presence of the nonlinear conversion medium inside the lasing cavity can also introduce challenges, such as complicating the laser's tunability and stability \cite{burdVECSELSystemsGeneration2016,hartkeExperimentalStudyOutput2008,kimAnalyticalModelIntracavity2008,calvezSemiconductorDiskLasers2009}. These challenges can compromise the use of intra-cavity frequency-converted VECSELs in applications exploiting the manipulation of accurate, and precise, energy transitions.

\begin{figure}[t]
\centering
\includegraphics[width=\linewidth]{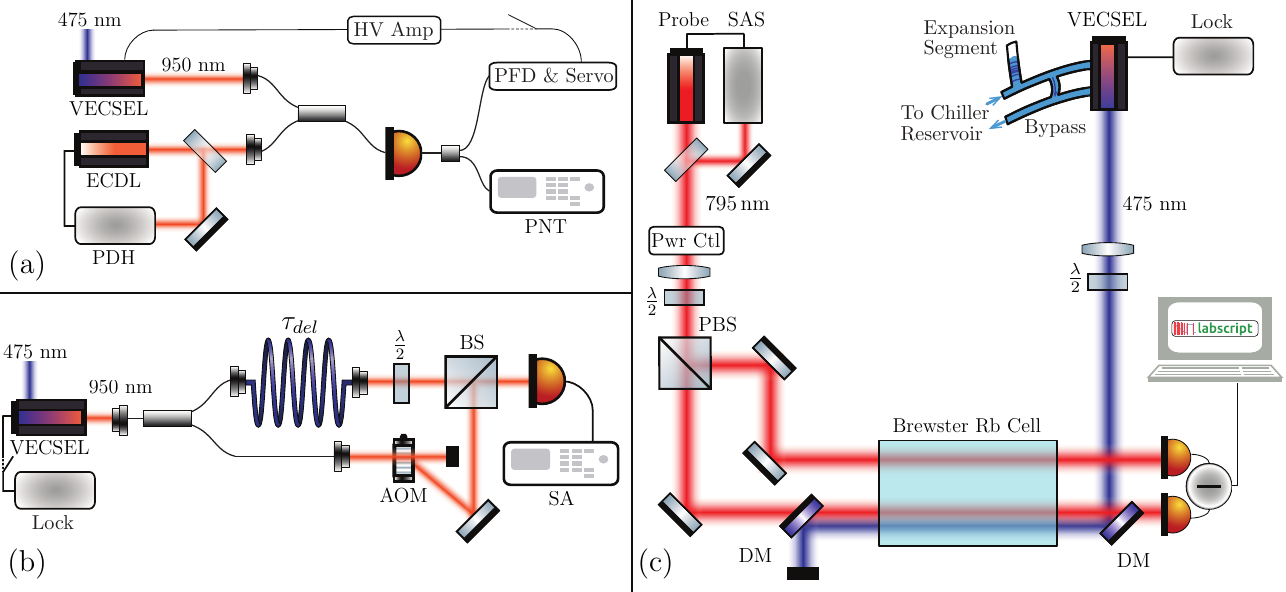}
\caption{Apparatus drawings for the three measurements performed in this work.
(a) Direct frequency noise analysis of the optical beat note between the ECDL reference and the VECSEL fundamental.
(b) Delayed Self-Heterodyne measurement.
(c) Rydberg-EIT spectroscopy measurement, incorporating the bypass and expansion segment modifications to the water chiller lines.
Abbreviations: (P)BS: (polarizing) beam splitter, SA: spectrum analyzer, PNT: phase noise tester, PDH: Pound-Drever-Hall frequency locking scheme, PFD: phase-frequency detector, HV Amp: high voltage amplifier, Lock: beat note lock to stabilized ECDL shown in (a), SAS: saturated absorption spectroscopy frequency locking scheme, Pwr Ctl: power control via a liquid crystal amplitude stabilizer, $\lambda/2$: half-waveplate, DM: dichroic mirror.
}
\label{fig:apparatus}
\end{figure}

The intra-cavity doubled VECSEL system we use and characterize in this work is a commercial prototype VALO SHG SF system produced by Vexlum Ltd\bibnote{This and all other references to commercial devices do not constitute an endorsement by the US Government or the Army Research Laboratory. They are provided in the interest of completeness and reproducibility.}.
It comprises an optically-pumped semiconductor gain medium inside an approximately 10\,cm long V-shaped lasing cavity containing a birefringent filter and etalon to enforce single-mode operation along with a nonlinear crystal for frequency doubling of the gain medium's emission. The gain medium, which also acts as one mirror of the cavity, is an epitaxially grown multi-layer semiconductor structure containing GaInAs quantum wells designed to emit fundamental (pre-doubled) light near 960\,nm under high pump irradiance.
Non-resonant optical-pumping of the gain medium allows the use of any pump laser with photon energy greater than the semiconductor band-gap energy.
Using the 808\,nm multimode diode bar from the prototype, VECSELs emitting at wavelengths up to 2\,$\upmu$m can be engineered by suitable changes to the semiconductor gain medium. For a more detailed description of the laser operation: see Ref.~\cite{tinsleyWattlevelBlueLight2021} that describes the gain medium, and Refs.~\cite{kantolaHighefficiency20Yellow2014,burdVECSELSystemsGeneration2016} that describe the intracavity-doubled VECSEL design.

In the current prototype, up to 14.7\,W of 808\,nm pump power is used to produce lasing at the fundamental wavelength that is tunable between approximately 932\,nm and 962\,nm. The tuning range for high-efficiency frequency doubling is limited to approximately 474.34\,nm to 481.22\,nm (948\,nm to 962\,nm in the fundamental) by phase matching of the lasing cavity with the intra-cavity nonlinear crystal. Coarse frequency tuning is accomplished by manually changing the rotation angle of a birefringent filter inside the lasing cavity, as well as the temperatures of various internal components (e.g. gain medium, filter, doubling crystal, etalon). One lasing-cavity mirror is mounted to a piezo-electric (PZT) actuator for fine frequency adjustments and stabilization via external feedback. The actuator's tuning sensitivity is $30$\,MHz/V and its lowest resonance is at 100\,kHz.
Given the present lack of other feed-forward tuning elements, the mode-hop-free tuning of the PZT typically ranges between 1-2\,GHz.

The laser's peak output power is 1.51\,W at 477\,nm with 14.7\,W of pump power, giving an optical-to-optical power efficiency of nearly $10\%$. Because of the inclusion of the doubling medium inside the lasing cavity,
it is challenging to obtain a smooth gain profile for the doubled output as phase-matching of the doubling crystal to the lasing cavity mode must be continuously maintained as frequencies are changed.Furthermore, due to an air-filled external cavity with high circulating power, this design is sensitive to atmospheric water absorption lines within the laser's fundamental wavelength operating range. This sensitivity can be mitigated by flowing gaseous dry nitrogen through the laser housing or sealing the laser housing with desiccant. The wavelength used for Rydberg-atom EIT presented in this work is approximately 475.238\,nm. At this wavelength, 13.7\,W of pump power produces peak visible light output of 690\,mW. Flowing nitrogen though the laser housing, thereby mitigating water absorption, increases the visible output to 780\,mW. For the rest of this manuscript, we use the system without flowing nitrogen to avoid instabilities arising from the sensitivity to mechanical fluctuations, as described next.

Mechanical perturbations are a major factor that influence the prototype laser's frequency stability, as has been noted for similar systems \cite{burdVECSELSystemsGeneration2016,zhang23wattSinglefrequencyVerticalexternalcavity2014,kasparLinewidthNarrowingPower2013}.
This has implications for the immediate environment in which the laser is operated. However, mechanical isolation and a sound-damping enclosure are simple steps (beyond those used in this work) that can provide protection against external influences. A more challenging manifestation of this sensitivity is chiller-induced frequency-noise caused by coolant moving through both the laser housing and the internal block to which the gain medium/mirror and its thermo-electric cooler (TEC) are attached. Similar findings have been reported in other VECSEL systems, illustrating a potential area for engineering improvements \cite{burdVECSELSystemsQuantum2020,laurain15SingleFrequency2014}. To characterize the chiller-induced fluctuations, we monitor the transmission of the laser through a high-finesse Fabry-Perot cavity after applying known electro-optically modulated sidebands. With high water flow level (3\,L/min), we observe fluctuations of approximately 15\,MHz root-mean-square (rms) in a 500\,$\upmu$s measurement time. Such fluctuations initially precluded locking this laser to a high-finesse Fabry-Perot cavity using the Pound Drever-Hall (PDH) technique. To limit the flow rate through the laser head, we add an external bypass line and T-branched, air-filled expansion segment (see Fig.~\ref{fig:apparatus}(c)). These coolant-line modifications reduce the frequency instability to less than 20\,kHz rms in a 500\,$\upmu$s measurement time, sufficient for locking to a number of quality frequency references. Chiller-induced fluctuations are likely endemic to many VECSELs that employ water cooling of the gain medium (which is one mirror of the extended cavity), and improvements that limit fluctuations due to water flow would be broadly beneficial. However, sensitivity to ambient acoustic noise (e.g. conversational voices) remains. For this reason, stabilization methods with a large capture range are favorable, such as those derived from the heterodyne beat (beat-note) between two lasers. We implement such a locking technique for all data the presented in this manuscript.

\section{Laser noise characterization}\label{sec:laserNoiseChar}

In semiconductor lasers, a simplified model for the power spectrum of the frequency noise $S_\nu$ includes two components: a frequency-independent (white) noise and a noise that scales as $1/f$ (pink noise) \cite{ludvigsenLaserLinewidthMeasurements1998}. We perform multiple tests to quantify these noise levels and generally characterize the frequency noise of the VECSEL. In Sections \ref{sec:DSH2.0} and \ref{sec:DSH10.7} we use the delayed self-heterodyne (DSH) technique, where the laser is beat against a time-delayed copy of itself, to measure both the intrinsic laser linewidth due to white noise only and that including $1/f$ components. In Section \ref{sec:BN} we compare the VECSEL to a stabilized reference laser via analysis of the beat note between the two lasers to quantify the $1/f$ noise that dominates at low frequencies. We perform both the DSH and beat note measurements using the VECSEL's fundamental light, and derive the corresponding values of the fundamental's laser linewidth. In the case of ideal frequency doubling, the visible output light's linewidth is twice that of the fundamental's.

These measurements include data while the device was free-running (unlocked), and while stabilized (locked).
We stabilize the VECSEL using an offset phase-locked servo to control a beat note between the fundamental ($\lambda \sim 950$\,nm) light from the laser and that of a commercial External Cavity Diode Laser (ECDL).
Feedback to the VECSEL is limited to that via the PZT, and the feedback loop bandwidth is approximately 10\,kHz. The ECDL reference laser is stabilized to an ultra-low expansion glass (ULE) optical cavity (10\,cm long, finesse greater than 10k) via the PDH method. Details of the performance of this lock are described in Section \ref{sec:BN}.

\subsection{Delayed self-heterodyne: 2.0\,km delay}
\label{sec:DSH2.0}

The delayed self-heterodyne technique is useful for determining the spectral characteristics of a single laser system \cite{ludvigsenLaserLinewidthMeasurements1998,okoshiNovelMethodHigh1980}.
By splitting light from a single laser source into two paths, and introducing a time delay in one prior to recombination, the original output can be interfered with a copy of itself. The interference spectrum contains information about the source's spectral purity, and depends on the relative timescales of the delay ($\tau_{del}$) and laser coherence ($\tau_{decoh}$).

We implement the DSH technique using fundamental light from a monitor port on the VECSEL, arranged to interfere in a Mach-Zehnder configuration, see Fig.~\ref{fig:apparatus}(b). The light is coupled into an optical fiber with an integrated beamsplitter. One path is delayed by including either 2\,km or 10.7\,km of telecommunications-wavelength fiber. The availability of nearly 100\,mW of light from the monitor port helps overcome significant attenuation in the fiber. The two paths exit collimating optics into free-space, where the non-delayed branch is shifted by $f_{AOM}=160$\,MHz with an acousto-optic modulator (AOM) resulting in a beat signal far away from low-frequency technical noise sources in the detection system. Finally, the beams are overlapped onto a 1.5\,GHz bandwidth photodiode using a nonpolarizing beamsplitter. Waveplates after the delaying fiber enable compensation for polarization state drifts that optimizes the interferometer signal. The photodiode output is recorded with a swept spectrum analyzer.

Assuming the laser exhibits only white frequency noise, an analytic form for the DSH power spectrum at variable delay times relative to the laser's coherence time, can be derived using the autocorrelation of the total laser field at the detector \cite{ludvigsenLaserLinewidthMeasurements1998,okoshiNovelMethodHigh1980,richterLinewidthDeterminationSelfheterodyne1986,zhangGeneralRelationFrequency2020,wangUltranarrowlinewidthMeasurementUtilizing2020}. In this framework, the power spectrum $S(\delta,\Delta)$ is represented as a combination of a Lorentzian shape and oscillations representing coherent interference that decay exponentially for larger detunings ($\Delta/2\pi=f-f_{AOM}$) of the Fourier frequency $f$ with respect to the carrier.
\begin{equation}\label{eq:whiteDSH}
    S(\delta,\Delta)=S_1S_2
\end{equation}
where
\begin{equation}
    S_1=\frac{A\delta}{\delta^2+\Delta^2}
\end{equation}
\begin{equation}\label{eq:coherentBeats}
    S_2=1-e^{-\delta \tau_{del}}\cdot \left[\cos(\Delta \tau_{del})+\delta \tau_{del} \,\text{sinc}(\Delta \tau_{del}) \right ]
\end{equation}
An additional resonant delta function term representing perfect coherence for $\tau_{del}\ll\tau_{decoh}$ is omitted in Eq.~\ref{eq:whiteDSH}, and $A$ is an overall scale factor. The oscillations' amplitude and frequency depends on the relative magnitudes of $\tau_{del}$ and $\delta$, the laser's Lorentzian linewidth due to white noise. If $\tau_{del}\lesssim\tau_{decoh}$, these oscillations are prominent and can be fit to Eq.~\ref{eq:whiteDSH} to extract the precise values for the delay time and the laser's Lorentzian linewidth due to white noise \cite{ludvigsenLaserLinewidthMeasurements1998}. As $\tau_{del}$ is increased, the visibility of these oscillations is suppressed. In the limit of $\tau_{del}\gg\tau_{decoh}$, Eq.~\ref{eq:whiteDSH} reduces to a single Lorentzian peak with a width twice that of the laser linewidth $\delta$.

\begin{figure}[tb]
\centering
\includegraphics[width=0.7\linewidth]{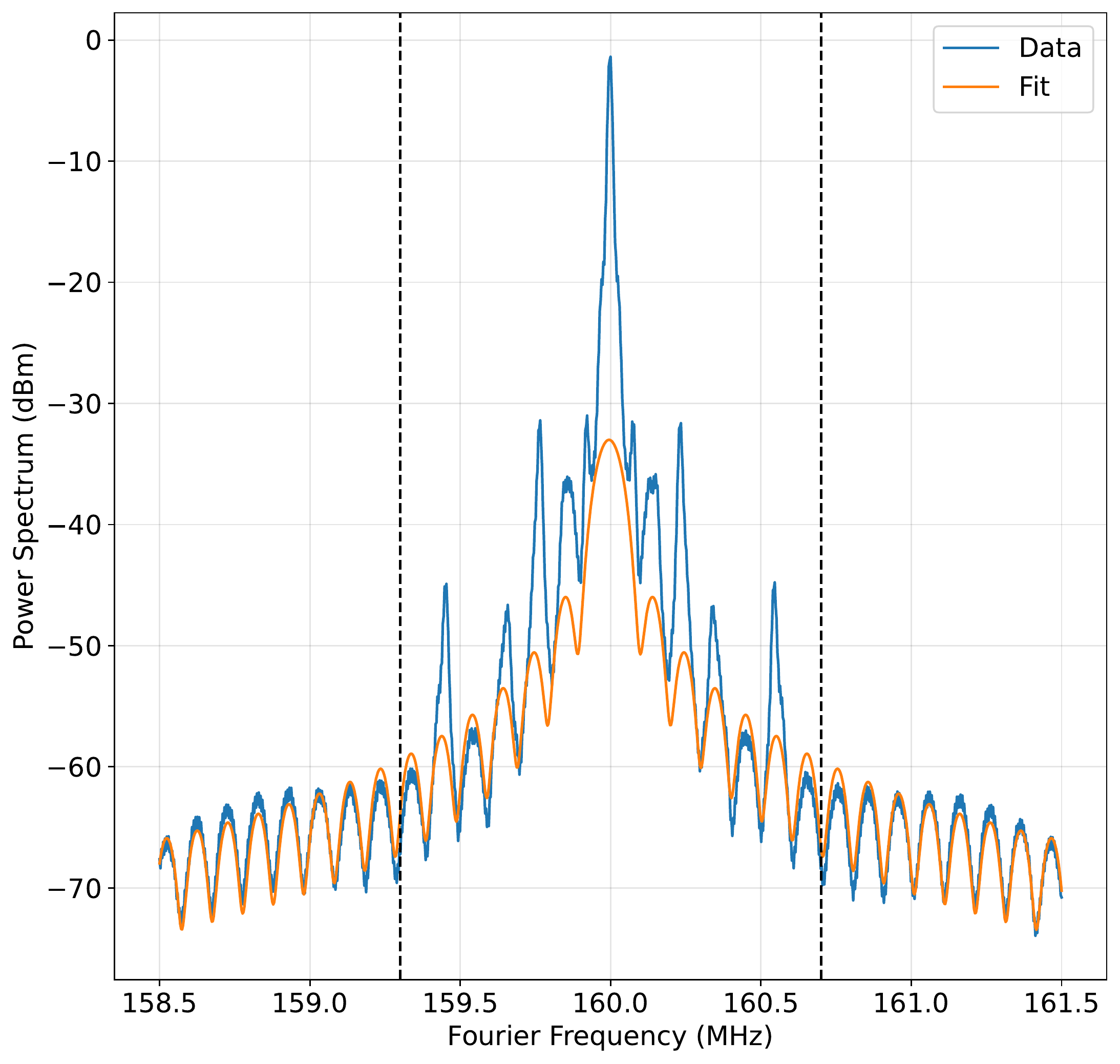}
\caption{Delayed self-heterodyne spectra of the VECSEL, while locked, with a 2\,km long fiber delay line ($\tau_{del}=9.8$\,$\upmu$s). Data with $|f-f_{AOM}|\geq 0.7$\,MHz (as designated by the vertical dashed lines) are fit to Eq.~\ref{eq:whiteDSH}. The results indicates a laser linewidth, accounting for white noise sources only, of $\delta =2\pi\times5.3(2)$\,kHz. The spectrum analyzer's resolution bandwidth was 8.2\,kHz.}
\label{fig:lorzFit}
\end{figure}

For a 2\,km delay line ($\tau_{del}=9.8$\,$\upmu$s), the locked VECSEL's DSH power spectrum displays oscillations that are described by Eq.~\ref{eq:whiteDSH}, as shown in Fig.~\ref{fig:lorzFit}. This is similar to results reported using the technique for characterization of direct-emission VECSELs \cite{rantamaki6WSingleFrequency2012}. At smaller detunings $\Delta$ from the carrier, however, the laser's DSH spectra display structure that is not captured by the model. This structure is indicative of other technical noise sources, possibly including the effect of pump light residual amplitude noise (RIN) on the gain medium's index of refraction \cite{laurainOpticalNoiseStabilized2014,liuNoiseInvestigationDualFrequency2018,myaraNoisePropertiesNIR2013,kasparLinewidthNarrowingPower2013}. Therefore, we fit only the data in the wings of Fig.~\ref{fig:lorzFit} to Eq.~\ref{eq:whiteDSH}, excluding 1.4\,MHz about $f_{AOM}$ as indicated by the dashed vertical lines. Locking the laser during this measurement suppresses contributions to the fit from $1/f$ noise (see Fig.~\ref{fig:PNT_BN_phasenoise}), helping highlight those originating from only white noise sources. From this, the laser's linewidth (accounting for white noise sources only) is determined to be $\delta=2\pi\times5.3(2)$\,kHz. This fit also enables an accurate determination of $\tau_{del}$, and allows us to extrapolate $\tau_{del}$ for longer fiber lengths. For the data of Figures \ref{fig:lorzFit} and \ref{fig:normSpectra}, the spectrum analyzer's resolution bandwidth (RBW) was 8.2\,kHz.
Each curve represents the average of 1000 traces acquired over 10 seconds, removing the calibrated spectrum analyzer's noise floor at -77\,dBm.

\subsection{Delayed self-heterodyne: 10.7\,km delay}
\label{sec:DSH10.7}

\begin{figure}[bt]
\centering
\includegraphics[width=0.7\linewidth]{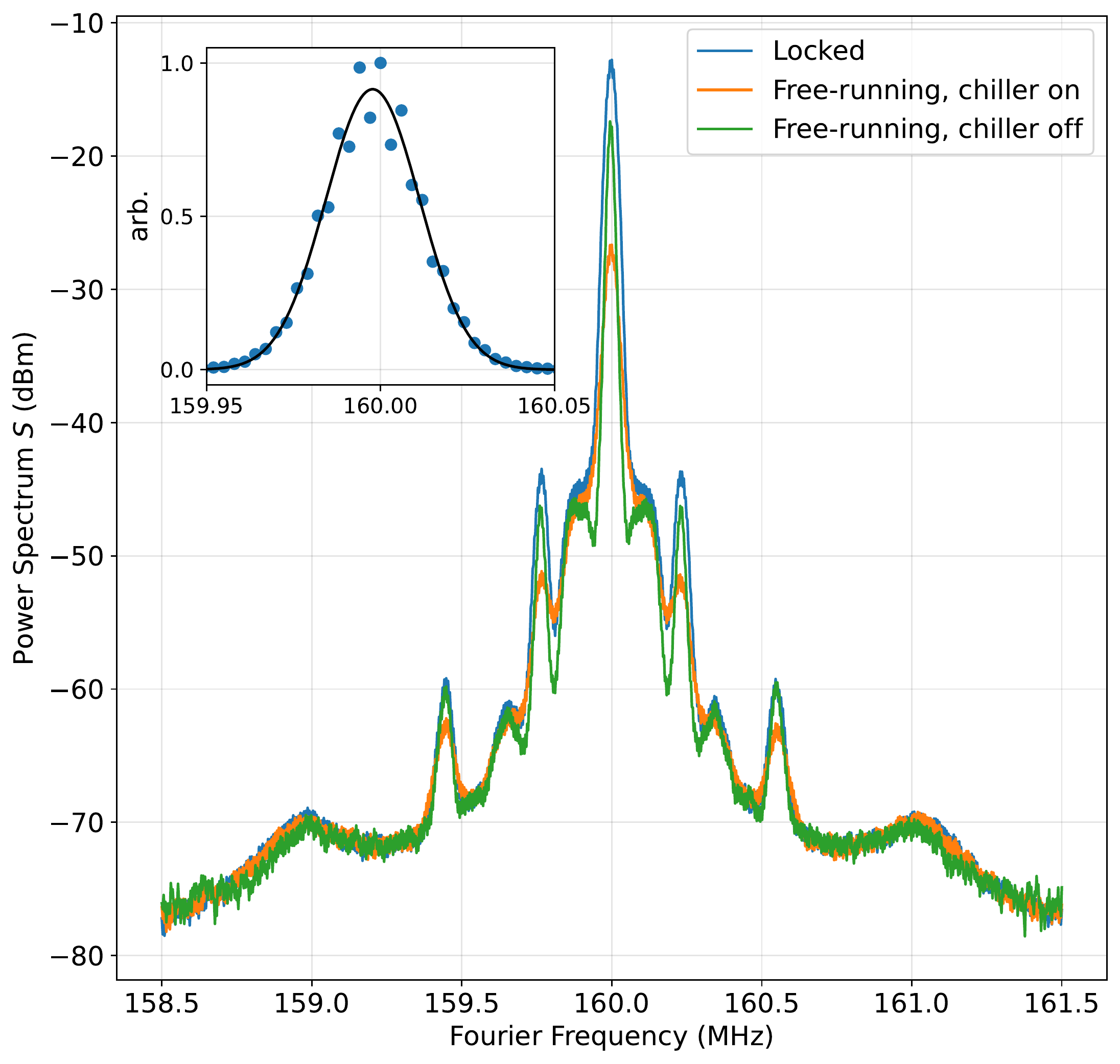}
\caption{Delayed self-heterodyne spectra of the VECSEL, with a 10.7\,km long fiber delay line ($\tau_{del}=53$\,$\upmu$s). 
Inset shows a Gaussian fit to the locked dataset central feature in normalized linear units. The corresponding gaussian linewidth is 23(2)\,kHz. The spectrum analyzer's resolution bandwidth was 8.2\,kHz}
\label{fig:normSpectra}
\end{figure}

We make similar DSH measurements using a 10.7\,km delay line ($\tau_{del}=53$\,$\upmu$s) with the VECSEL locked, and free running. The coherent oscillations, visible at shorter delay times, are suppressed with this longer delay indicating the laser's coherence time is less than the delay time. As a result, the technical noise is made more apparent (see Fig.~\ref{fig:normSpectra}). When the laser is free-running, the chiller can be turned off briefly for the duration of the measurement to assess its impact on the spectra. We observe that doing so reduces the laser's frequency fluctuations, sharpening the structure of the power spectrum signal in a way that is similar to locking the laser.

Because of changes in the detected optical power due to drifts and/or turning off the chiller, we scale the free-running data in Fig.~\ref{fig:normSpectra} by the ratio $P_{L}/P$ such that the data in the wings of the spectra overlap. $P_{L}$ is the the total power (integrated power spectrum) in the locked trace and $P$ is the same quantity for the corresponding free-running data.
The spectrum scale factors with the chiller on and off, are 0.5\,dB and 5.6\,dB, respectively.

The effect of $1/f$ frequency noise from the laser becomes more apparent with increasingly long ($\tau_{del}\gg\tau_{decoh}$) delay lines, that further broadens the DSH spectra. Such broadening can be approximated as a Gaussian shape contribution to the spectra for small carrier offset frequencies, having a linewidth that is dependent on $\tau_{del}$ \cite{mercerFrequencyNoiseEffects1991}. The inset of Fig.~\ref{fig:normSpectra} shows a Gaussian fit to that figure's locked dataset, in linear units normalized to the peak power. The Gaussian fit FWHM of the DSH spectrum, $\gamma=32(3)$\,kHz, corresponds to the laser's $1/f$ linewidth of $\Gamma_G/2\pi=\gamma/\sqrt{2}=23(2)$\,kHz \cite{mercerFrequencyNoiseEffects1991}, as measured with $\tau_{del}=53$\,$\upmu$s. The corresponding laser linewidths in the free-running cases are 28\,kHz and 14\,kHz with the chiller on, and off, respectively.

\subsection{Beat-note frequency deviations}
\label{sec:BN}

\begin{figure}[tb]
\centering
\includegraphics[width=0.7\linewidth]{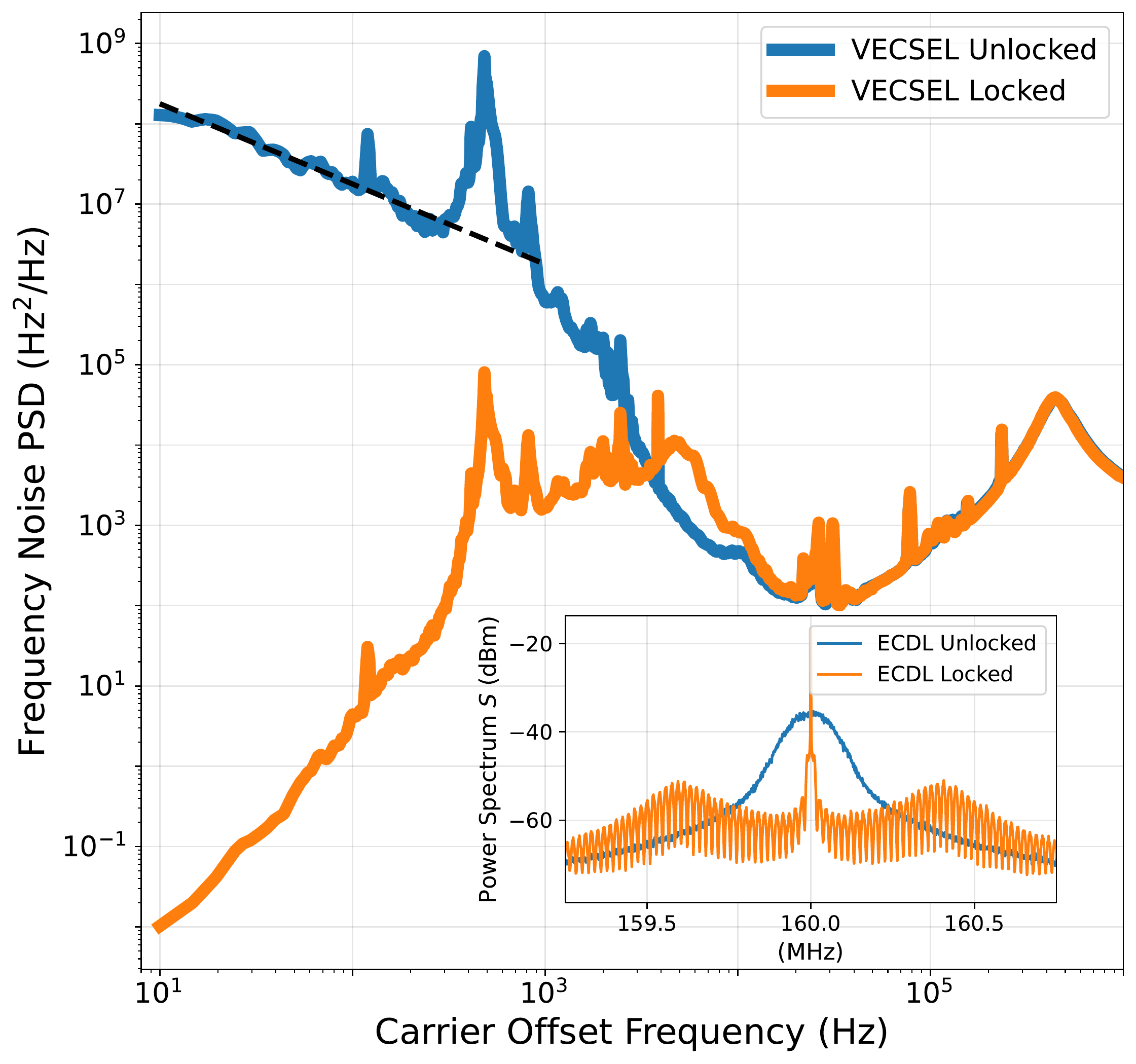}
\caption{Frequency noise power spectral density (PSD) of a beat note between VECSEL fundamental light and a commercial ECDL. In all cases the ECDL remains locked via PDH to a high-finesse optical cavity. The orange (blue) trace corresponds to the VECSEL servo being locked (unlocked) to the ECDL with the setup shown in Fig.~\ref{fig:apparatus}(a).  The black dashed line represents the $1/f$ frequency noise dominated portion of the free-running laser, and has the form $k/f$ with $k=1.78\times10^9$\,Hz$^3$/Hz. The PDH locking servo is responsible for the ``servo bump'' at about 440\,kHz. The inset shows delayed self-heterodyne spectra of the reference ECDL when locked, or unlocked, to the optical cavity. The locking servo's narrowing effect, and ``bump'', are evident.}
\label{fig:PNT_BN_phasenoise}
\end{figure}

In addition to the DSH technique, we also analyze the VECSEL's frequency deviations by beating its fundamental light against that from a commercial ECDL acting as a reference. This reference is locked via the PDH method to a high-finesse optical cavity. We can characterize the reference laser's output via 10.7\,km delay line DSH measurements when locked and unlocked, as shown in inset of Fig.~\ref{fig:PNT_BN_phasenoise}. Fitting the unlocked spectrum to a Lorentzian, we determine the reference laser's free-running Lorentzian linewidth to be 15(1)\,kHz. We fit the locked DSH spectrum within $\pm150$\,kHz of the carrier to Eq.~\ref{eq:whiteDSH}. From this measurement, we extract an equivalent Lorentzian linewidth of less than three kHz for the locked reference laser. While the lock does narrow the coherent central feature significantly (and provides absolute frequency stability at long time scales), a detrimental ``servo bump'' at 440\,kHz is also visible as increased amplitude coherent oscillations in the locked DSH spectrum.

Next, we form a beat note by interfering the VECSEL's fundamental light with that of the reference laser on a $1.5$\,GHz bandwidth, fiber-coupled, photodetector. We characterize the VECSEL-reference beat note using a phase noise tester\bibnote{A Berkeley Nucleonics Corporation Model 7340} referenced to a rubidium frequency standard. The beat note can also be used to phase-lock\bibnote{A Vescent D2-135 implements the phase-frequency detector and PID servo.} the VECSEL and reference laser.
The frequency-noise power spectral density (PSD) with the VECSEL locked and unlocked is presented in Fig.~\ref{fig:PNT_BN_phasenoise}. 
The beat note lock suppresses frequency deviations of the beat frequency at slow time scales by up to 100\,dB. The maximum achievable feedback bandwidth was approximately 10\,kHz, limited by the piezo resonance near 100\,kHz.
A black dashed line overlapping the unlocked data represents the $1/f$ frequency noise dominated portion of the free-running laser, and has the form $k/f$ with $k=1.78\times10^9$\,Hz$^3$/Hz. The magnitude of this noise is not readily apparent in the DSH measurements described above due to the finite delay times used; analysis against a stable reference is necessary to quantify it. The PSD at frequencies above approximately 40\,kHz is dominated by noise on the ECDL reference due to the PDH locking servo. Because of the high offset noise in the reference, the VECSEL's noise at similar frequencies is obscured and we must rely on the DSH measurements above to quantify it.

\subsection{Frequency stability summary}

Taken together, beat note and DSH measurements quantify this VECSEL's spectral characteristics in a manner consistent with those of similar prototypes, including the direct-emission variety \cite{rosenerUmSemiconductorDisk2011,rantamakiWatt56Um2013,laurainMultiwattpowerHighlycoherentCompact2010,tourrencLowfrequencyFMnoiseinducedLineshape2005,vicianiLineshapeVerticalCavity2002}. Due to the the laser's technical noise, and varying methods of specifying spectral purity depending on the noise source, timescale involved, and measurement parameters (e.g. DSH delay fiber length), there is no single definitive linewidth value that completely describes this system.

The prototype displays a Lorentzian linewidth due to white frequency noise of 5.32\,kHz, as measured when locked using a short delay (Fig.~\ref{fig:lorzFit}) to remove contributions from technical noise. Conversely, significant $1/f$ technical noise (28.1\,kHz, $\tau_{del}=52.71\, \upmu$s) is evident when measured free-running with a long delay (Fig.~\ref{fig:normSpectra}) that maximizes sensitivity to low-frequency deviations. Such low-frequency noise introduces long-term instability that can be mitigated by common frequency-stabilization techniques, up to the available feedback bandwidth. In the case of this VECSEL, that bandwidth is limited by the combined PZT/servo response to $\sim 10$\,kHz. The nontrivial structure in the noise spectrum may preclude some applications, but is sufficiently suppressed with respect to the carrier to be appropriate for Rydberg-EIT in thermal vapors. For example, in the free-running, chiller-on, dataset of Fig.~\ref{fig:normSpectra}, the leading-order $\pm 230$\,kHz noise contribution has a relative magnitude of -25\,dBc.

\section{Rydberg-EIT spectroscopy}

Electromagnetically Induced Transparency is a coherent nonlinear phenomena typically observed in quantum systems consisting of three discrete energy states. Interference between excitation channels modifies the medium's complex susceptibility, increasing the transmission of a weak probing laser, due to the presence of a strong coupling laser (see inset of Fig.\ref{fig:EITspectra}) \cite{fleischhauerElectromagneticallyInducedTransparency2005}.
In the case of Rydberg-EIT, one of the three states is a Rydberg state with high principal quantum number $n\gtrsim 10$
\cite{mohapatraCoherentOpticalDetection2007}. Such states are interesting due to their exaggerated properties that scale with $n$, and are an active field of research with applications in quantum computing and sensing \cite{meyerWaveguideCoupledRydbergSpectrum2021,meyerAssessmentRydbergAtoms2020,adamsRydbergAtomQuantum2020,saffmanQuantumComputingAtomic2016}. 

Rydberg-EIT in a thermal ensemble of atoms results in a narrow spectroscopic feature ($\lesssim10$\,MHz) that has proven particularly useful for RF and microwave electric field sensing \cite{sedlacekMicrowaveElectrometryRydberg2012}. The transition wavelength that couples to these Rydberg states in alkali atoms is typically in the visible spectrum, and because the dipole moment of that transition is small, significant optical powers are necessary. Light sources that satisfy these requirements have historically been difficult to obtain, making VECSELs a promising candidate for use in such coherent excitation schemes. The gain medium can be engineered to lase at a variety of wavelengths, outputting high power into a single spatial mode while being broadly tunable across many atomic resonances. 
In short, Rydberg-EIT represents a measurement that requires and demonstrates the advantageous qualities of intra-cavity frequency-doubled VECSELs for other quantum science applications.

\subsection{Spectroscopy setup}

The ladder-type Rydberg-EIT system consists of the 5\,S$_{1/2}$, 5\,P$_{1/2}$, and 41\,$D_{3/2}$ states in $^{85}$Rb as shown in the inset of Fig.~\ref{fig:EITspectra}. The experimental layout is shown in Fig.~\ref{fig:apparatus}(c). The atomic sample consists of natural isotopic abundance Rb vapor at room-temperature enclosed in a 13\,cm long glass cell. The cell's endcap windows are positioned at Brewster's angle with respect to the incident beams to minimize reflective losses. Such losses are under 1\% per window, with metallic Rb deposits contributing to excess loss of transmitted light.

The 474.34\,nm to 481.22\,nm operating range of the VECSEL described above covers excitation from the $5P_{1/2}$ state to the n\,D$_{3/2}$ Rydberg states with principal quantum numbers $n=21-62$.
The selected Rydberg state 41\,$D_{3/2}$ is excited at 630824.918\,GHz ($\sim 475.235$\,nm). The dipole moment of this optical transition is $1.547\times10^{-2}\,ea_0$, where $e$ is the elementary charge and $a_0$ is the Bohr radius.
The Rydberg excitation beam (blue in Fig.~\ref{fig:EITspectra}) is called the ``coupling'' beam. This vertically-polarized beam is free-space propagated through expansion optics, and at the location of the vapor cell, has a nearly circular intensity profile with $1/e^2$ radius of 1.25\,mm. We obtain 690\,mW of power in the coupling laser, incident on the atomic sample, corresponding to a Rabi frequency of $\Omega_c=2\pi\times2.88$\,MHz. This and other properties of the Rydberg states and transitions are calculated using the ARC python package \cite{sibalicARCOpensourceLibrary2017}.

The probe beam is derived from an external cavity diode laser (ECDL) resonant with the $^{85}$Rb D1 $F=3\leftrightarrow F'=3$ transition ($\lambda_p \sim 794.979$\,nm). $F$ and $F'$ are the initial and final total orbital angular momentum quantum numbers of the hyperfine structure manifold that is unresolved in this room-temperature sample. The probe laser's frequency is referenced to a separate Distributed Bragg Reflector (DBR) laser that is stabilized to the $^{85}$Rb 5\,S$_{1/2}$,$F=2 \leftrightarrow 5P_{1/2}$, $F'=3$ transition via saturated absorption spectroscopy performed in a separate vapor cell. The beat note between the DBR reference and probe lasers is stabilized to ensure the probe light frequency is fixed and resonant with the $^{85}$Rb D1 $F=3\leftrightarrow F'=3$ transition.

After exiting fiber-collimation optics, the probe beam passes through a liquid-crystal power stabilization unit. Next, the beam is expanded and split into two paths using a waveplate and polarizing beamsplitter. The two beams are vertically polarized, and propagate parallel to each other through the cell, 1\,cm apart. One probe beam counterpropagates against an overlapping coupling beam before the colors are separated with a dichroic mirror. At the vapor cell, the overlapped probe beam has a circular intensity profile with a $1/e^2$ radius of 1.2\,mm. With 6\,$\upmu$W of power incident on the cell, the corresponding Rabi frequency is $\Omega_p=2\pi\times 0.988$\,MHz. We use the two probe beams to implement a balanced detection scheme, as shown in Fig.~\ref{fig:apparatus}(c). This allows cancellation of non-EIT contributions to the signal.

The coupling beam's frequency is stabilized via the beat note lock described in section three: centered on the
$^{85}$Rb 5\,P$_{1/2} \leftrightarrow 41\,$D$_{3/2}$ Rydberg transition. To perform spectroscopy, the probe beam is held on resonance ($\Delta_p=\omega_p - \omega_0=0$) while the coupling beam's frequency ($\Delta_c = \omega_c - \omega_1$) is swept by ramping the beat note setpoint reference tone using the linear sweep accumulator of an AD9959 direct digital synthesizer (DDS) clocked by a 100\,MHz low-noise frequency reference. Experiment control and data acquisition is performed using the labscript suite \cite{starkeyScriptedControlSystem2013}.

\subsection{Results and discussion}

\begin{figure}[t]
\centering
\includegraphics[width=0.7\linewidth]{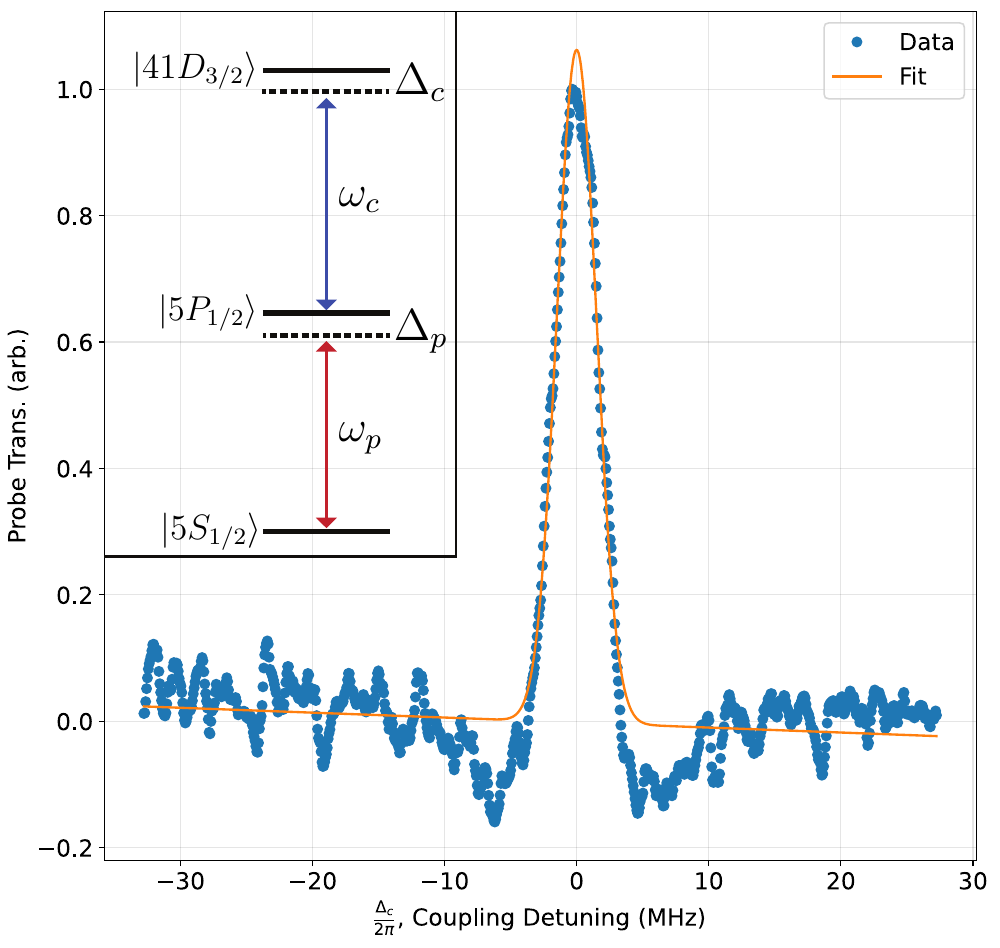}
\caption{EIT spectra and Gaussian fit for a single sweep of the coupling beam detuning $\Delta_c$ with the probe beam frequency fixed at single-photon resonance ($\Delta_p = \omega_p - \omega_0=0$). Transmission values are normalized to the peak value. The fit's full-width half-maximum is 3.5\,MHz.
Inset depicts the $^{85}$Rb ladder-type EIT excitation pathway used in this work.}
\label{fig:EITspectra}
\end{figure}

The EIT spectroscopy results are shown in Fig.~\ref{fig:EITspectra}.
The data are collected in a single sweep, lasting 3.6\,ms. 
Signals from the photodetector are low-pass ($f_{lp}=10$\,kHz) and notch ($f_n=60$\,Hz) filtered to remove residual spurious noise from the acquisition electronics (that sample at 250\,kHz). The full-width-half-maximum of the EIT transmission peak is 3.5\,MHz, measured using a Gaussian fit function (yellow in Fig. \ref{fig:EITspectra}). This is, to the best of our knowledge, the narrowest reported atomic spectroscopy using a frequency-doubled VECSEL.
It is also consistent with narrow Rydberg-EIT linewidths reported in the literature that range from 2-10\,MHz \cite{mohapatraCoherentOpticalDetection2007,kumarAtomBasedSensingWeak2017,meyerDigitalCommunicationRydberg2018,jingAtomicSuperheterodyneReceiver2020,prajapatiEnhancementElectromagneticallyInduced2021}.
Our observed linewidth is not limited by power broadening, confirmed by identical linewidths for reduced optical powers. Rydberg-EIT in thermal vapors via two-photon excitation is unable to achieve narrower linewidths due to Doppler-averaging effects in mis-matched ladder systems \cite{mohapatraCoherentOpticalDetection2007,ripkaApplicationdrivenProblemsRydberg2021}. There is no indication from the measured EIT linewidth that the frequency noise of the VECSEL's doubled output significantly differs from that of the fundamental, which was characterized above.

Obtaining such a narrow Rydberg-EIT feature and favorable signal-to-noise ratio, particularly for a 3.6\,ms measurement window in a thermal ensemble, demonstrates the utility of this laser for quantum sensing applications.
Narrow Rydberg-EIT that avoids power broadening necessitates using smaller Rabi frequencies that also reduce the transparency peak amplitude.
In this situation, one typically must increase the measurement time to maintain sufficient signal-to-noise.
For example, the narrowest reported two-photon Rydberg-EIT linewidth of 1.7\,MHz used a one second integration time \cite{kumarAtomBasedSensingWeak2017}.
An alternative to long integration times is to increase the beam size.
The VECSEL's high output power and stability was crucial to enable this for our system.
The high output power allows for increasing beam size while maintaining a sufficiently large Rabi frequency on the coupling transition ($\Omega_c\propto\sqrt{P_c}/r$).
Large beams are advantageous for probing more atoms \cite{coxQuantumLimitedAtomicReceiver2018,meyerOptimalAtomicQuantum2021}, and for maintaining higher probe power to overcome detector noise limitations, while limiting power broadening. Larger beams also reduce transit broadening that stems from the finite interaction time as atoms transit through the laser beam \cite{sagleMeasurementModellingIntensity1996}. The VECSEL's spatial mode quality also leads to efficient overlap with the probing field profile, requiring fewer beam shaping components. 

\section{Conclusion}

We have investigated a VECSEL that employs an engineered quantum well gain medium and intra-cavity frequency doubling to produce laser light between 475 to 481\,nm.
The laser exhibits high output power, high spectral purity, broad tunability, and a circular transverse spatial mode, all in the UV to visible part of the spectrum where such qualities are difficult to obtain simultaneously.
These qualities are of especially high demand in quantum science applications that involve ionic, neutral atom, or solid-state platforms where transitions commonly occur at these wavelengths and significant power is required.
We have characterized this system's stability and frequency characteristics using delayed self-heterodyne and direct beat note comparison to a narrow reference ECDL. We find linewidths due to white noise only, and including $1/f$ noise while stabilized, of 5.3(2)\,kHz and 23(2)\,kHz, respectively.

Looking forward, there are areas of laser engineering that would further the capabilities of this platform. Improved thermal management to reduce the effect of vibrations from the water chiller, improvements to limit acoustic sensitivity, engineering to remove the spurious linewidth structure, a faster PZT feedback path, the addition of pump current modulation to improve feedback bandwidth, and active pump intensity stabilization are all areas undergoing active development. However, current performance of the prototype is sufficient for many applications. We demonstrated the use of this laser system in one such application by measuring Rydberg-EIT spectroscopy in a thermal vapor, as is commonly done in Rydberg RF electrometry. We observe a Rydberg-EIT transmission window with a FWHM of 3.5\,MHz; the narrowest atomic spectroscopy feature observed using an intra-cavity doubled VECSEL. This measurement was done in 3.6\,ms; faster than similar measurements reported in the literature. The laser's high output power and optimal spatial output mode were critical to observing this narrow linewidth.
By demonstrating MHz-level atomic atomic spectroscopy and characterizing the laser's frequency noise with two independent and complementary techniques, we have highlighted the potential for intra-cavity frequency-doubled VECSELs to have a significant impact on many atom-optical quantum technologies.

\begin{backmatter}

\bmsection{Acknowledgments} The authors thank James Cahill for use of the fiber delay-line spools, and Fredrik Fatemi for thoughtful discussion. J.-P.P. acknowledges the support of the Jenny and Antti Wihuri Foundation, the Walter Ahlström Foundation, and the Finnish Foundation for Technology Promotion. 

\bmsection{Disclosures} JPP: Vexlum Ltd. (I,E,P), EK: Vexlum Ltd. (I,E,P)

\bmsection{Data availability} 
Data underlying the results presented in this paper may be obtained from the authors upon reasonable request.

\end{backmatter}

\bibliography{VECSEL}

\end{document}